\documentclass{amsart}

\usepackage{amsbsy,amssymb,amscd,amsfonts,latexsym,amstext,delarray,
amsmath,graphicx} 
\input xypic
\usepackage{color}

\newtheorem{thm}{Theorem}[section]
\newtheorem{prop}[thm]{Proposition}
\newtheorem{cor}[thm]{Corollary}
\newtheorem{lem}[thm]{Lemma}
\newtheorem{defn}[thm]{Definition}

\newtheorem{ex}[thm]{Example}
\newtheorem{ques}[thm]{Question}

\numberwithin{equation}{section}

\def\bL{{\mathbb L}}

\def\bT{{\mathbb T}}

\def\A{{\mathbb A}}
\def\C{{\mathbb C}}
\def\F{{\mathbb F}}

\def\N{{\mathbb N}}
\renewcommand{\P}{{\mathbb P}}
\def\Q{{\mathbb Q}}
\def\Z{{\mathbb Z}}
\def\R{{\mathbb R}}
\def\K{{\mathbb K}}

\def\cM{{\mathcal M}}

\def\cS{{\mathcal S}}
\def\cT{{\mathcal T}}

\def\cV{{\mathcal V}}

\def\cX{{\mathcal X}}

\def\cZ{{\mathcal Z}}

\def\fp{{\mathfrak{p}}}
\def\fq{{\mathfrak {q}}}
\def\fA{{\mathfrak{A}}}
\def\fZ{{\mathfrak{Z}}}

\title{Arithmetic of Potts model hypersurfaces}
\author{Matilde Marcolli and Jessica Su}
\address{Mathematics Department, Mail Code 253-37, Caltech, 1200 E.~California Blvd. Pasadena, CA 91125, USA}
\email{matilde@caltech.edu} 
\email{jessicas@caltech.edu}

\begin{document}
\maketitle

\begin{abstract}
We consider Potts model hypersurfaces defined by the multivariate Tutte
polynomial of graphs (Potts model partition function). We focus on the behavior of 
the number of points over finite fields for these hypersurfaces, in 
comparison with the graph hypersurfaces of 
perturbative quantum field theory defined by the Kirchhoff graph polynomial.
We give a very simple example of the failure
of the ``fibration condition" in the dependence of the Grothendieck class
on the number of spin states and of the polynomial countability condition for these
Potts model hypersurfaces.
We then show that a period computation, formally similar to the parametric
Feynman integrals of quantum field theory, arises  by considering certain
thermodynamic averages. One can show that these evaluate to combinations
of multiple zeta values for Potts models on polygon polymer chains, while
silicate tetrahedral chains provide a candidate for a possible occurrence
of non-mixed Tate periods. 
\end{abstract}

\section{Introduction}

A lot of attention was devoted  in recent years to the intriguing
occurrence of periods of algebraic varieties and motives in the context of
perturbative quantum field theory and residues of Feynman integrals.
The question initially arose from numerical observations of  Broadhurst
and Kreimer \cite{brokr} on the occurrence of multiple zeta values in the 
computation of Feynman diagrams of massless scalar field theories. 
Multiple zeta values were conjectured to be the periods of mixed Tate 
motives, a result recently proved by Francis Brown \cite{brown1}.  Moreover,
the parametric formulation of Feynman integral (see \cite{bjdrell}, \cite{itzub})
exhibits the residues of
Feynman diagrams explicitly as periods of an algebraic variety (the 
graph hypersurface complement), at least modulo divergences and
renormalization. It was then natural to expect that the occurrence of
multiple zeta values would be an indication of the fact that the graph
hypersurfaces are mixed Tate motives. This was conjectured by 
Kontsevich, in a substantially equivalent statement predicting that these
hypersurfaces would be polynomially countable. This means that
their counting of points over finite fields would depend polynomially
on the order of the field. At first this conjecture was confirmed (see
\cite{stanley}, \cite{stembridge}) for all graphs with up to twelve 
edges, but it was later disproved by a very elegant and general
result of Belkale and Brosnan \cite{belbro}, which shows that, contrary
to expectations, the classes of the graph hypersurfaces span a
suitable localization of the Grothendieck ring of varieties, hence they
can be arbitrarily far from the mixed Tate case as motives. For a 
discussion of the difference between the behavior of the classes
in the localized or in the ordinary Grothendieck ring see also \cite{aluffi2}. 
The original cases observed numerically in \cite{brokr} were later
confirmed to be periods of mixed Tate motives in \cite{brown2},
\cite{brown3}, \cite{schnetz2} and other cases in which the parametric
Feynman integral turns out to be a period of a mixed Tate motive
were analyzed in \cite{aluffi3}, \cite{bek}, \cite{blokr}, \cite{doryn2}, \cite{schnetz}, 
\cite{schnetz2}, and others. 

\medskip

While the result of Belkale and Brosnan 
\cite{belbro} that disproved the Kontsevich conjecture is very general
and elegant, it is not immediately constructive, in the sense that it
does not exhibit an explicit graph that fails to satisfy the polynomial
countability property. The first such explicit example was identified
by Doryn in \cite{doryn}, and other examples were then constructed
in \cite{brownschnetz}. There are interesting conceptual differences
between these examples, which point to two different forms in which
the polynomial countability property can fail. One, as in the case
of \cite{doryn}, where the dependence in the order of the field
fails to satisfy {\em the same polynomial} function at all primes,
and another more drastic one, where an actual {\em non-polynomial 
function} appears.

\medskip

In the recent paper \cite{aluffi}, it was shown that some of the techniques
employed in the study of the classes of the graph hypersurfaces in the 
Grothendieck ring of varieties can be extended to another type of
physical model, this time not in quantum field theory but in statistical
mechanics. In fact, the partition function for the Potts model for a spin
system on a graph with $q$ possible states is given by a combinatorial
polynomial associated to the graph, the multivariate Tutte polynomial, 
see \cite{sokal}. This is closely related to the Kirchhoff polynomial that 
defines the graph hypersurfaces in quantum field theory. In the context
of Potts models, one is less directly interested in the periods of the
hypersurface complement, and more in describing how the hypersurface
behaves when over families of graphs that approximate some infinite
graph. In fact, zeros of the partition function correspond to physical
phase transitions, if they happen for non-negative real values of the
parameters, while they can be thought of as some kind of non-physical
``virtual" phase transitions otherwise. 
While (at least in the ferromagnetic case) the Potts model
on a finite graph does not have phase transitions, which can
only appear in the limit of an infinite graph, in the antiferromagnetic
case there are possible phase transitions also over a finite graph 
\cite{sokal}. The sets of real zeros of the multivariate Tutte polynomials 
have been extensively studied in the Potts model literature, most
notably by Jackson and Sokal \cite{jacsok}. In \cite{aluffi} it is shown
that one can estimate how the complexity of this set of zeros grows
in certain families of graphs that approach infinite graphs, in terms
of classes in the Grothendieck ring of the varieties defined by the
multivariate Tutte polynomials, the Potts model hypersurfaces.
These are analyzed using a deletion-contraction type formula
similar to the one obtained in \cite{aluffi4} for the graph hypersurfaces
of quantum field theory.

\medskip

In the present paper, continuing to draw on the analogy between the
graph hypersurfaces and the Potts model hypersurfaces, 
$\cZ_{G,q}$, for a graph $G$ and for a number $q$ of spin states,
we compare the behavior of the counting of points over finite fields for these varieties.
We first consider the ``fibration condition" discussed in \cite{aluffi}, that is, the
question of whether the Grothendieck class of $\cZ_{G,q}$ depends 
on $q$ outside of the special values $q=0$ and $q=1$. 
This question is motivated by an 
observation made in \cite{aluffi}, where it was shown that, 
for certain classes of graphs (such as polymer chains), the classes of the 
hypersurfaces $\cZ_{G,q}$ behave as if the hypersurfaces were 
a locally trivial fibration away from those two special values. We show here
that the fibration condition already fails for the tetrahedron graph $K4$, and
we argue that one can use a Monte Carlo method, similar to 
the one developed in \cite{montecarlo}, \cite{montecarlo2}, to
test the failure of the fibration condition for Potts model hypersurfaces
of more complicated graphs.  We then look at the polynomial
countability question for the Potts model hypersurfaces. Naturally,
one expects that failures of polynomial countability will appear
for much smaller and combinatorially simpler graphs than in the
cases of the graph hypersurfaces of \cite{brownschnetz} and \cite{doryn}.
Indeed, this is the case.  A trivial kind of failure of 
polynomial countability comes from the primes that divide the
number of spin states $q$, but even after reformulating the 
question by taking into account this trivial failure, we see that,
again, the tetrahedron graph $K4$ already provides an example
where a more interesting failure of polynomial countability is possible.

\medskip

We discuss the role of periods in this statistical mechanical setting,
where we interpret them in terms of certain thermodynamic averages. 
Unlike the more difficult case of quantum field theories, where infrared
divergences arise from the intersection between the locus of integration
in the parameteric Feynman integral and the graph hypersurfaces, in
the (ferromagnetic) Potts model case the hypersurface does not intersect
the simplex over which the integration is performed, hence we can
directly interpret the resulting integral as a period, without the
serious additional complication of blowups performing regularization
and renormalization, as one finds (\cite{bek}, \cite{blokr}) in the
quantum field theory case. 

\medskip

The main purpose of this paper is pedagogical. It is aimed at
presenting this approach to Potts models through 
the algebro-geometric and motivic properties of the hypersurfaces
defined by the multivariate Tutte polynomials, in terms of very
concrete and simple examples, which, we hope, can serve as an
illustration of the general methodology.

\medskip

\section{Graph hypersurfaces and Potts model hypersurfaces}

To fix notation, in the following we denote by $G=G(V,E)$ a finite graph,
with vertex set $V$ and edge set $E$. We also write $V=V(G)$
and $E=E(G)$. 

\smallskip
\subsection{Quantum field theory and graph hypersurfaces}

We first recall briefly the definition of the graph hypersurfaces of 
perturbative quantum field theory.

\begin{defn}\label{graphpolyndef}
Let $G=G(V,E)$ be a graph with $n=\# V$. 
Assign an edge weight (a variable) $t_e$
to each edge $e\in E$. 
The Kirchhoff graph polynomial of $G$ is of the form
\begin{equation}\label{graphpoly}
\Psi_G(t_1,\ldots, t_n)= \sum_{T\subset G} \prod_{e\notin E(T)} t_e ,
\end{equation}
where the sum is over all the spanning trees $T$ (or maximal spanning
forests in the multi-connected case).
The graph hypersurface $X_G$ is the hypersurface in $\A^n$ defined by
the vanishing of the graph polynomial
\begin{equation}\label{graphhyp}
\cX_G =\{ t=(t_1,\ldots,t_n)\in \A^n \,|\, \Psi_G(t)=0 \}.
\end{equation}
\end{defn}

Since the graph polynomial is homogeneous of degree $b_1(G)$, the
affine graph hypersurface is the affine cone over a projective graph
hypersurface in $\P^{n-1}$.

\smallskip

The role of the graph hypersurfaces in perturbative quantum field theory
is explained by the following result (see \cite{bjdrell}, \cite{itzub}, \cite{mar}).
The Feynman integral associated to the graph $G$ is given by
\begin{equation}\label{Feynman}
 U(G)= \int_{\sigma_n} \frac{P_G(p,t)^{-n+D(n-1)/2}}{\Psi_G(t)^{n(-1+D/2)}} \omega_n, 
\end{equation} 
where $D$ is the spacetime dimension, $P_G$ is the second graph polynomial (whose
explicit form we do not recall here), with $p$ the external momenta, $\sigma_n$
the unit simplex, and $\omega_n$ the volume form.
Here the Feynman integral refers to the residue (after removing a divergent
Gamma factor), written in momentum
space and in the Feynman parametric form, for the case of 
a massless scalar field theory (see the references listed above for 
more details).  Modulo important issues of divergence and renormalization, the
integral \eqref{Feynman} is then a period of the graph hypersurface
complement. 

\medskip

\subsection{Statistical mechanics and Potts model hypersurfaces}

The analogous object  we will be discussing in the statistical mechanics setting 
is the Potts model partition function. 

\smallskip

To avoid notational ambiguity, we will write $\fp$ for an integer prime
and $\fq=\fp^n$ for a prime power, so that we will be using the notation 
$\F_{\fq}$ is the finite field with
$\fq$ elements. We reserve the notation $q$ for the variable in the Potts
model partition function that denotes the number of spin states. 

\smallskip

\begin{defn}\label{defmultiTutte}
Let $G=G(V,E)$ be a graph with $n=\# V$, and with edge weights $t=(t_e)_{e\in V}$.
The multivariate Tutte polynomial of $G$ is 
\begin{equation}\label{multiTutte}
T_G(q,t) = \sum_{A \subseteq E} q^{k(A)} \prod_{e \in A} t_e ,
\end{equation} 
where the sum is over the subgraphs of $G$ that span all vertices (that is, subgraphs $G'$
with $V(G')=V(G)$ and $E(G')=A\subset E$), and where $k(A)$ is the number of connected components in each subgraph, and $q$ is an indeterminate.
\end{defn}

To see the relation to Potts models in statistical mechanics, we consider a set $\fA$ of
cardinality $q$ (here assumed to be a non-negative integer) of possible spin states,
and a statistical system of spins assigned to the vertices of a graph $G$. A state
of the spin system if an assignment of a spin state $\sigma_v$ to each vertex. The energy of 
a state is the sum over all edges of a quantity 
$$ H(e) = \left\{ \begin{array}{cc} 0 & \sigma_v \neq \sigma_w \\
- J_e & \sigma_v=\sigma_w,
\end{array} \right. $$
where $\partial(e)=\{ v, w\}$, and where $J_e$ 
is a fixed value, with $J_e \geq 0$ in the ferromagnetic case
and $-\infty \leq J_e \leq 0$ in the antiferromagnetic case. The weight edges
are then related to the energies $J_e$ by the relation 
$$ t_e = e^{\beta J_e} -1, $$
where $\beta$ is (up to the Boltzmann constant) an inverse temperature. The
partition function of the Potts model statistical system on $G=G(V,E)$ is then given by
the expression
\begin{equation}\label{PottsZ}
Z_G(q,t) = \sum_{\sigma: V\to \fA} \prod_{e\in E} (1+ t_e \delta_{\sigma_v, \sigma_w}),
\end{equation}
where the sum is over all states (that is, all maps of vertices to spin states), 
$\delta$ is the Kronecker delta, and $\partial(e)=\{ v, w\}$.
The relation of the multivariate Tutte polynomial \eqref{multiTutte} 
to statistical mechanics comes from a famous
result of Fortuin--Kasteleyn \cite{forkas}, which shows that the partition function
of the Potts model is a restriction of the multivariate Tutte polynomial
\begin{equation}
Z_G(q,t) = T_G(q,t) |_{q\in \N, t=e^{\beta J}-1 \in \R} .
\end{equation}

\medskip

The Potts model hypersurface is the locus of zeros defined by the multivariate Tutte polynomial
\begin{equation}\label{ZGhyp}
\cZ_G =\{ (q,t)\in \A^{n+1} \,|\, Z_G(q,t)=0 \}.
\end{equation}
Notice that, unlike the Kirchhoff graph polynomial, the multivariate Tutte polynomial is
not homogeneous, so this is an affine hypersurface in $\A^{n+1}$, where $n=\# E(G)$,
but not a projective hypersurface. One also considers the
Potts model hypersurface for fixed $q$,
\begin{equation}\label{qZGhyp}
\cZ_{G,q} =\{ t \in \A^n \,|\, Z_G(q,t)=0 \}.
\end{equation}

\medskip

The relation between the Potts model hypersurfaces and the graph hypersurfaces
of quantum field theory lies in the fact (see \cite{aluffi}, \cite{sokal}) that the Kirchhoff
graph polynomial $\Psi_G$, or rather its equivalent form
\begin{equation}\label{PhiG}
\Phi_G(t)= \sum_{T\subset G} \prod_{e\in E(T)} t_e,
\end{equation}
which is related to $\Psi_G$ by dividing by $\prod_{e\in E} t_e$ and applying 
the Cremona transformation $t_e \mapsto 1/t_e$ to all the edge variables, is
obtained by normalizing the $\tilde T_G(q,t)=q^{-k(G)} T_G(q,t)$, evaluating 
at $q=0$ and taking the homogeneous piece of lowest degree. This relation has
a more geometric interpretation in terms of tangent cones, as in Lemma 2.7
of \cite{aluffi}. This relation is the motivation for extending techniques developed
in the quantum field theory context for studying the geometry of the graph
hypersurfaces to the setting of Potts models.

\subsection{Counting points and classes in the Grothendieck ring}

Let $\cV_\K$ denote the category of (quasi-projective)
algebraic varieties over a field
$\K$. The Grothendieck ring of varieties $K_0(\cV_\K)$ is the free
abelian group generated
by the isomorphic classes $[X]$ of varieties, modulo the relations
$$ [X] = [Y] + [X\smallsetminus Y] $$
for closed embeddings $Y\subset X$ of subvarieties. The product
operation that gives the ring structure is given by $[X]\cdot [Y]=[X \times Y]$.

\smallskip

For an affine hypersurface $X \subset \A^n$, we will use in the following the notation $\{ X \}$ for
the class in $K_0(\cV_\K)$ of the hypersurface complement 
\begin{equation}\label{complclass}
\{ X \} :=[ \A^n \smallsetminus X ] =[\A^n]-[X] =\bL^n - [X],
\end{equation}
where $\bL=[\A^1]$ is the Lefschetz motive, the class of the affine line.

\smallskip

Similarly, one can define a Grothendieck ring $K_0(\cV_\Z)$ for
arithmetic varieties defined over $\Z$. While the relations are
formally the same, the classes $[X]$ in this case denote isomorphism
classes as varieties over $\Z$. We will sometime write only
$K_0(\cV)$, without specifying the field or ring of definition, 
when it should be clear from the context.

\smallskip

An additive invariant is a map $\chi : \cV_\K \to R$, with values in a commutative
ring $R$, that satisfies $\chi(X)=\chi(Y)$ if $X$ and $Y$ are isomorphic, and
$\chi(X)=\chi(Y)+\chi(X\smallsetminus Y)$, for closed embeddings $Y\subset X$.  
Moreover, one requires that $\chi(X\times Y)=\chi(X)\chi(Y)$.
An additive invariant determines and is determined by a ring homomorphism
$\chi : K_0(\cV_\K) \to R$. For varieties in $\cV_\C$, the topological Euler
characteristic is an additive invariant, and the properties of being additive 
under decompositions $X=Y\cup (X\smallsetminus Y)$ and multiplicative
under products are its defining properties, hence one can consider the
class $[X]$ in the Grothendieck ring as a {\em universal Euler characteristic}
of the variety $X$, see \cite{bitt}. 

\smallskip

Recall that for every prime number $\fp$ and every $n\geq 1$ there
is a unique field extension $\F_\fq$ of degree $n$ of $\F_\fp$, with $\fq=\fp^n$. 
In the case of varieties $X$ defined over  a finite field $\F_\fp$. One denotes then by
$X (\F_{\fp^m})$ the set of $\F_{\fp^m}$-points of $X$, for $m\geq 1$. One then sets
\begin{equation}\label{NqX}
N_{\fp^m}(X) = \# X (\F_{\fp^m}), 
\end{equation}
the number of $\F_{\fp^m}$-points of $X$. Then it is not hard to see that
$N_{\fp^m}(X)$ is an additive invariant of $X$, hence it factors through
$K_0(\cV_{\F_{\fp}})$, so that we can write $N_{\fp^m}([X])$ as a
function of the class $[X]$ in the Grothendieck ring.

\smallskip

In the following, we will be considering varieties $X$ defined over $\Z$. 
In this case, we can either regard them as complex varieties
by embedding $\Q \subset \C$, or reduce modulo $\fp$, at the various
primes, and obtain varieties $X_\fp$ defined over $\F_\fp$, for which one can
consider the counting of points $N_{\fp^m}(X)$. 

\smallskip

We are interested in the behavior of the number of points $N_{\fq}(X)$
as a function of $\fq$. We recall the following terminology (see \cite{belbro},
\cite{stanley}, \cite{stembridge}).

\begin{defn}\label{polycount}
Let $X$ be a variety over $\Z$ with reduction $X_\fq$ over $\F_\fq$, for $\fq=\fp^m$. 
Then $X$ is polynomially countable if $N_{\fq}(X_\fq)$ is a polynomial function of $\fq$.
\end{defn}

\smallskip

Because the counting of points over finite fields is an additive invariant, which
factors through the Grothendieck ring of varieties, one can use polynomial
countability as a test for the motivic nature of the hypersurfaces, as follows.

\smallskip

If $\bL=[\A^1]$ is the Lefschetz motive in $K_0(\cV)$, 
then $\Z[\bL] \subset K_0(\cV)$ is the subring of the Tate motives. More precisely,
one should view the Tate motives as $\Z[\bL,\bL^{-1}] \subset K_0(\cM)$, for
$\cM$ the category of pure Chow motives, but for our purposes it suffices to
work with $K_0(\cV)$. There is, anyway, a homomorphism $\chi_{mot}: K_0(\cV) \to
K_0(\cM)$ defined by the additive invariant of \cite{gilsou}. For varieties that are
not smooth and projective, one leaves the category of pure motives and one
needs to consider objects in a more complicated (triangulated) category of
mixed motives, which contains a (triangulated) subcategory of mixed Tate
motives. However, even varieties that are singular still define classes
in the Grothendieck ring, and the $\Z[\bL]$ part of the Grothendieck ring
contains the classes of the varieties whose motive is a mixed Tate motive.

\smallskip

Then, if a variety $X$ defined over $\Z$ is a mixed Tate motive, it has class 
$[X]\in \Z[\bL]$ in $K_0(\cV_\Z)$. Away from up to finitely many primes $\fp$ 
(where some bad reduction phenomenon can occur), it then follows that 
$N_{\fp^m}(X_{\fp})$ is a polynomial in $\fp^m$, by the 
fact that the counting $N_{\fp^m}(X_\fp)$ factors through the Grothendieck 
ring $K_0(\cV_{\F_{\fp}})$ and that $N_{\fp^m}(\A^1)=\fp^m$.
Thus, if a variety $X$ does not have the property of being {\em polynomially 
countable at all but finitely many primes}, 
it follows that it is not a mixed Tate motive.

\smallskip

In fact, polynomial countability at all but finitely many primes is 
conjecturally equivalent to the motive of the
variety being mixed Tate: the Tate conjecture predicts that determining 
$N_{\fp}(X_\fp)$ for almost all primes $\fp$ would determine the motive of $X$, see
\cite{andre}.

\section{The fibration condition}

In \cite{aluffi} a specific condition for the classes in the Grothendieck
ring of the Potts model hypersurfaces is identified, called the ``fibration condition",
according to which the the Grothendieck class $\{ \cZ_G \}$ of the complement of the hypersurface
$\cZ_G$ behaves as one would expect in the case of a fibration on the locus $q \neq 0, 1$. 
This means that the class $\{ \cZ_{G,q} \}$ is independent of $q$, for all $q\neq 0,1$, and that
\begin{equation}\label{fibcond}
 \{ \cZ_G \} = (\bT -1) \cdot \{ \cZ_{G,q\neq 0,1} \} + \bT^{\# E(G)}. 
\end{equation} 

This condition is satisfied by the families of graphs considered in \cite{aluffi}, but
the general question of whether it holds for more general graphs was not addressed
in that paper, nor was the question of whether, when the condition on the
Grothendieck classes is satisfied, the variety $\cZ_G$ is really a 
locally trivial fibration over the locus $q\neq 0,1$.

Here we show that, in fact, very simple
examples of graphs already do not satisfy the fibration condition and the 
number of points over $\F_\fp$ of $\cZ_{G,q}$ has a nontrivial dependence 
on $q\neq 0,1$.

\begin{prop}\label{K4nofibration}
The graph $K4$ does not satisfy the fibration condition.
\end{prop}

\proof One can see this by a direct computation of the number of points over
$\F_\fp$ for different values of $q$. The following table illustrates the
failure of the fibration condition.

\begin{center}
    \begin{tabular}{|c|c||c|}
    \hline
$\fp$ & $q$ & $\#    \cZ_{G,q}( \F_\fp )$ \\ \hline\hline
11 & 0 &  1771561  \\ \hline
11 & 1 & 771561 \\ \hline
11 & 2 &  173799  \\ \hline
11 & 3 & 173183 \\ \hline
11 & 4 & 173821 \\ \hline
11 & 5 & 173513 \\ \hline
11 & 6 & 174151 \\ \hline
11 & 7 & 173227 \\ \hline
11 & 8 & 173447 \\ \hline
11 & 9 & 173579  \\ \hline
11 & 10 & 173799  \\ \hline
        \end{tabular}
\end{center}

\endproof

By comparison, polygons are the simplest example of graphs
that do satisfy the fibration condition of \cite{aluffi}.
In fact, by Proposition 5.2 of \cite{aluffi} we know that
for a polygon $C_{m+1}$ with $m+1$ sides, one has
\begin{equation}\label{polygonclass}
\{ \cZ_{C_{m+1},q\neq 0,1} \} = \bT^{m+1} + \bT (\bT^m -(\bT-1)^m) + \frac{(\bT-1)^m -(-1)^m}{\bT}, 
\end{equation}
with $\bT =\bL-1$, and it is therefore independent of $q\neq 0,1$.
Comparing this expression with Proposition 5.1 of \cite{aluffi}, one sees
that indeed \eqref{fibcond} holds for the polygons.

\medskip

\section{Monte Carlo method for counting points}

Monte Carlo methods for counting points of varieties over finite fields
were introduced in \cite{montecarlo2} and \cite{montecarlo}. 
We employ here the algorithm described in \cite{montecarlo2}
and we show how it compares with the deterministic counting
of points, on a sufficiently simple example of a graph where
both computations can be performed. We use again the example
of the tetrahedron graph $K_4$, which we already discussed 
in the previous section.

\smallskip

The following tables provides an explicit comparison between the
deterministic counting and the Monte Carlo method (after 10000, 40000, 
and 100000 trials, respectively) for $K_4$, in the case $q=2$.

\begin{center}
    \begin{tabular}{|c||c|c|c|}
      \hline
$\fp$ & Monte Carlo & \% error & \% error bound \\ \hline\hline
3 & 413.9262 & 0.002242615 & 0.03363242 \\ \hline
5 & 4507.8125 & 0.013219263 & 0.047440384 \\ \hline
7 & 20670.9293 & -0.011007641 & 0.060059259  \\ \hline
11 & 179459.1293 & -0.004845872 & 0.079343222 \\ \hline
13 & 396763.6998 & -0.015535761 & 0.087605918 \\ \hline
17 & 1469977.952 & -0.015716002 & 0.101770275 \\ \hline
19 & 2399339.931 & -0.070218248 & 0.108087334 \\ \hline
23 & 6928079.605 & 0.045288885 & 0.119636754 \\ \hline\hline
3 & 412.1037 & -0.002170218 & 0.01681621 \\ \hline
5 & 4453.125 & 0.000927175 & 0.023720192 \\ \hline
7 & 21141.5253 & 0.011507837 & 0.03002963 \\ \hline
11 & 182825.0952 & 0.013819407 & 0.039671611 \\ \hline
13 & 404848.6049 & 0.004524793 & 0.043802959 \\ \hline
17 & 1503167.109 & 0.006507159 & 0.050885138 \\ \hline
19 & 2560472.073 & -0.007777023 & 0.054043667 \\ \hline
23 & 6661615.005 & 0.005085466 & 0.059818377 \\ \hline\hline
3 & 413.74395 & 0.001801332 & 0.010635505 \\ \hline
5 & 4477.5 & 0.006405934 & 0.015001967  \\ \hline
7 & 20710.92996 & -0.009093825 & 0.018992405 \\ \hline
11 & 183480.5728 & 0.017454225 & 0.02509053 \\ \hline
13 & 402024.9216 & -0.00248143 & 0.027703424 \\ \hline
17 & 1499908.538 & 0.004325248 & 0.032182587 \\ \hline
19 & 2532009.315 & -0.018806787 & 0.034180216 \\ \hline
23 & 6580195.266 & -0.007198912 & 0.037832464 \\ \hline
    \end{tabular}
\end{center}

\medskip

The values should be compared with the deterministic values reported
in the table in Proposition \ref{ZK4nonpol}.
The expected error bound is computed as in \cite{montecarlo2}, with
an error margin $\epsilon$ of  $\sqrt{4b (log(2/\delta))/N}$, 
where $b$ is the fraction of points that are roots, $N$ is the number of trials, and $1 - \delta$ is the probability of being within $\epsilon$ of the correct fraction. 

\smallskip

The Monte Carlo method is useful to test properties such as the fibration condition, when 
direct deterministic computations become intractable. In fact, when using the 
Monte Carlo method in cases with non-constant $q$-dependence, the deviance 
from a constant value is typically greater than what allowed by the error estimate.

\section{Failures of polynomial countability}

We now analyze failures of polynomial countability for the Potts model
hypersurfaces, which confirms the fact that these hypersurfaces are
much less likely to be mixed Tate motives than the graph hypersurfaces
of quantum field theory, although families of mixed Tate cases can
still be constructed, using the deletion--contraction relation derived in
\cite{aluffi}, in cases where a good recursive procedure is possible,
such as the polymer chains analyzed in \cite{aluffi}.

\smallskip

We use a variant of the algorithm used by
Stembridge \cite{stembridge} in the case of the graph hypersurfaces.
This can be adapted to our setting, because some general results
proved in \cite{stembridge} apply both to the graph polynomials
and the multivariate Tutte polynomials, in particular those
discussed in Lemma \ref{lemZdelcon} below.

\subsection{The special case $q=1$}

There is a very special case, which can be treated separately by direct
geometric reasoning, which is the case with $q=1$. Notice that this is
not a physically relevant case, since $q$ is the number of spin
states. 

\smallskip

When $q=1$, the multivariate Tutte polynomial assumes the very simple form
$$ Z_G(q,t)|_{q=1}=\sum_{A \subset E}\prod_{e \in A} t_e = \prod_{e \in E} (1 + t_e). $$
In fact, multiplying this out gives all possible combinations of the $t_e$'s, which 
correspond to all possible subsets $A$ of $E$.

\begin{lem}\label{q1count}
In the case $q=1$, the counting of points over $\F_\fp$ is given by
$$ N_{\fp}(\cZ_{G,q=1}) = \fp^{\# E(G)} - (\fp - 1)^{\# E(G)}. $$
\end{lem}

\proof 
The edge weight tuple satisfies the Tutte polynomial iff at least one of the edge weights $t_e$ is $-1$.  Therefore, we can get the number of roots by taking the total number of tuples and subtracting the number of tuples where none of the edge weights is $-1$.  If there are $\fp$ elements in the field and $m$ edges, the number of roots is $\fp^m - (\fp-1)^m$.  Therefore, the number of roots always depends polynomially on $\fp$ when $q = 1$.
\endproof

\smallskip

One can also see this directly from the class $[\cZ_{G,q=1}]$ in the Grothendieck
ring, as observed in \cite{aluffi}, the locus of zeros of the polynomial $\prod_{e \in E} (1 + t_e)$ is
isomorphic to the union of coordinate hyperplanes, whose complement in $\A^{\# E(G)}$
is a torus, whose class is $\bT^{\# E(G)}$. Thus, the class of the complement is
$\{ \cZ_{G,q=1} \}=\bT^{\# E(G)}$ and $[ \cZ_{G,q=1} ] =\bL^{\# E(G)} - \bT^{\# E(G)}$,
which immediately gives back the formula of Lemma \ref{q1count}, since
$N_{\fp}(\bL)=\fp$ and $N_{\fp}(\bT)=\fp-1$.

\medskip 
\subsection{A trivial failure for $q\neq 1$}

We now consider the case where $q\neq 1$ (we also assume $q\neq 0$).
In this case, we show that there is always a ``trivial" failure of polynomial
countability, which has to do with the primes that divide $q$. In fact, in the
way it is originally written, the multivariate Tutte polynomial has an overall
multiplicative factor of $q^{k(G)}$, with $k(G)$ the number of connected 
components of $G$. Thus, whenever $\fp$ is a prime that divides $q$,
the polynomial becomes identically zero in the reduction mod $\fp$. 

\smallskip

We start with the following simple observation.

\begin{lem}\label{onepoly}
If a set of points satisfies a single variable polynomial at infinitely many values but does not satisfy that polynomial at certain other values, the entire set of points cannot satisfy a single variable polynomial (even if it is a different polynomial).
\end{lem}

\proof Suppose given an infinite set $\cS$ with an infinite subset $\cT \subset \cS$,
such that the points of $\cT$ satisfy a polynomial $Q(x)$ of degree $m$ and the points
of $\cS$ satisfy a polynomial $P$ of degree $n$, then these two
polynomials must be the same. In fact,  both polynomials have to agree on
the infinite set $\cT$, hence their difference $R(x)=P(x)-Q(x)$ is a polynomial
with infinitely many zeros, that is, the trivial polynomial $R\equiv 0$.
\endproof

We then have the following general behavior.

\begin{cor}\label{Nnonpol}
For a fixed value $q\neq 0,1$ of the number of spin states, let $G$ be a graph 
for which the class $[\cZ_{G,q}]$ in $K_0(\cV_\Z)$ is in $\Z[\bL]$. Then the
counting function $N_{\fp}(\cZ_{G,q})$ is given by
\begin{equation}\label{NLq}
N_{\fp}(\cZ_{G,q}) =\left\{ \begin{array}{ll}  \fp^{\# E(G)}  &  \fp | q \\
P_q(\fp)    & \fp \not| 
q  \end{array} \right.
\end{equation}
where $[\cZ_{G,q}]=P_q(\bL)= a_0 + a_1 \bL + \cdots + a_N \bL^N$ in $\Z[\bL]$.
Thus, $N_{\fp}(\cZ_{G,q})$ is not a polynomial function.
\end{cor}

\proof If $q = 0$ in the field (that is, if $\fp$ is a divisor of $q$), everything is a root, so there are $\fp^m$ roots where $m$ is the number of edges.  If $q \neq 0$ ($\fp$ does not divide $q$), 
by the form $[\cZ_{G,q}]=P_q(\bL)= a_0 + a_1 \bL + \cdots + a_N \bL^N$ of the class in the 
Grothendieck ring we obtain that $N_{\fp}(\cZ_{G,q}) =P_q(\fp)$. The last
observation then follows from Lemma \ref{onepoly}.
\endproof

Notice that this type of failure of the polynomial countability, due only to the 
primes that divide the number $q$ of spin states, should be regarded as an
``accidental" and not a ``serious" failure of polynomial countability, in the sense 
that, for example, it does not really affect the nature of the classes in the
Grothendieck ring. 

\smallskip

We can see this explicitly in some simple examples. 
Consider first the case where $G=T$ is a tree. Then the multivariate
Tutte polynomial is of the form
\begin{equation}\label{treeTutte}
Z_T(q,t)= q \prod_{e \in E(T)} (q + t_e).
\end{equation}

\begin{ex}\label{treenonpol}
The counting function  $N_{\fp}(\cZ_{T_m,q})$ for a tree $T_m$ with $m$ edges is given by
\begin{equation}\label{Ntree}
N_{\fp}(\cZ_{T_m,q})=\left\{ \begin{array}{ll} \fp^m  &  \fp | q \\
\fp^m - (\fp - 1)^m &  \fp \not| q
\end{array} \right.
\end{equation}
\end{ex}

Another very simple explicit example is that of polygons,
for which the expression \eqref{polygonclass} for the class in the
Grothendieck ring gives the following.

\begin{ex}\label{polygonsnonpol}
The counting function  $N_{\fp}(\cZ_{C_{m+1},q})$ for a polygon $C_{m+1}$ is given
by $ \fp^{m+1}$ when $ \fp | q$, while when $\fp \not| q$, it is given by
\begin{equation}\label{polynon}
N_{\fp}(\cZ_{C_{m+1},q})= 
\fp^{m+1} -
( (\fp-1)^{m+1}   + (\fp-1) ((\fp-1)^m -(\fp-2)^m) + \frac{(\fp-2)^m -(-1)^m}{(\fp-1)}) .
\end{equation}
\end{ex}

Another way to see that 
$N_{\fp}(\cZ_{C_{m+1},q})$ is a polynomial in $\fp$, for all $\fp$ that do not divide $q$,
which does not use directly the expression  \eqref{polygonclass} for the class
in the Grothendieck ring, is to show, as in Stembridge \cite{stembridge} that a
partial deletion-contraction relation holds at the level of the counting functions.
This reflects the partial deletion-contraction relation at the level of classes in
the Grothendieck ring proved in \cite{aluffi}.

\medskip
\subsection{Probabilistic counting functions}

We recall the relevant notation from \cite{stembridge}. Given a set of
polynomials $\{ f_1,\ldots, f_k \}$ in $\Z[x_1,\ldots, x_m]$, one writes $\fZ[f_1,\ldots,,f_k]$ for
the probability that all the $f_i$ vanish at a uniformly randomly chosen $(x_1,\ldots,x_m)
\in \F_{\fq}$. Then, as a function of the field cardinality $\fq=\fp^r$, one has
\begin{equation}\label{Zfi}
\fZ[f_1,\ldots,,f_k] (\fq) = \fq^{-m} N_{\fq}(X_\fp),
\end{equation}  
where $X_\fp$ is the reduction mod $\fp$ of the variety defined over $\Z$ by the
polynomials $\{ f_1, \ldots, f_k\}$ and $N_\fq$ is the number of its $\F_\fq$-points,
for $\fq$ a power of $\fp$. Then by Proposition 2.1 of \cite{stembridge}, the
variety $X$ is polynomially countable if and only if $\fZ[f_1,\ldots,,f_k] (\fq)$ is a 
polynomial in $\fq^{-1}$. We will refer to $\fZ[f_1,\ldots,,f_k]$ as the ``probabilistic
counting function".

\smallskip

Then we have the following analog of Lemma 3.2 of \cite{stembridge}.

\begin{lem}\label{lemZdelcon}
Let $Z_{G,q}=Z_G(q,t)$ denote the multivariate Tutte polynomial, for fixed $q\neq 0,1$,
as a function of the edge weights $t=(t_e)$, and let $\fZ[ Z_{G,q} ]$ be the 
corresponding probabilistic counting function. Then this satisfies the relation
\begin{equation}\label{Zdelcon}
\fZ[Z_{G,q}] (\fp) = \fp^{-1} - \fp^{-1} \fZ[Z_{G/e,q}] + \fZ[Z_{G/e,q}, Z_{G \smallsetminus e,q}].
\end{equation}
\end{lem}

\proof
The multivariate Tutte polynomials satisfy a deletion-contraction relation
\begin{equation}\label{delconTutte}
Z_G(q,t) = Z_{G\smallsetminus e} (q, \hat t^{(e)}) + t_e \, Z_{G/e}(q,\hat t^{(e)}),
\end{equation}
for any edge $e$ (regardless of whether it is a regular edge, a bridge, or a looping edge),
with $\hat t^{(e)}$ denoting the set of edge weights with $t_e$ removed, and where
$G/e$ is the graph obtained by contracting $e$ in $G$ and $G\smallsetminus e$ the 
graph obtained by deleting $e$ in $G$.

Then the argument of Proposition 2.3 of \cite{stembridge} and the
deletion--contraction relation \eqref{delconTutte} show that one obtains the relation
\eqref{Zdelcon}.
\endproof

A more refined deletion-contraction relation of a similar nature to \eqref{Zdelcon} 
was proved by algebro-geometric methods in \cite{aluffi}, for classes in
the Grothendieck ring. Namely, for the classes of the hypersurface complements one has
\begin{equation}\label{K0delcon}
\{ \cZ_{G,q} \} = \bL \cdot \{ \cZ_{G/e,q} \cap \cZ_{G\smallsetminus e,q} \} - \{ \cZ_{G/e,q} \}.
\end{equation}

One sees then that one can also recover \eqref{Zdelcon} directly from \eqref{K0delcon}.
Indeed, we have from \eqref{K0delcon}, that (away from primes that divide $q$) the
counting function satisfies
$$ \begin{array}{rl}
N_{\fp}(\A^{\# E(G)}\smallsetminus \cZ_{G,q}))  = &
\fp  N_{\fp}(\A^{\# E(G)-1}\smallsetminus (\cZ_{G/e,q} \cap \cZ_{G\smallsetminus e,q}) )) \\
- &  N_{\fp}(\A^{\# E(G)-1}\smallsetminus  \cZ_{G/e,q})) , \end{array} $$
which gives
 $$ N_{\fp}( \cZ_{G,q})) = \fp   N_{\fp}(\cZ_{G/e,q} \cap \cZ_{G\smallsetminus e,q}) 
 +  \fp^{\# E(G)-1} - N_{\fp}(\cZ_{G/e,q}). $$
 Thus, one obtains, as expected,
$$ \fZ[Z_{G,q}] (\fp) = \fp^{-\# E(G)} N_{\fp}(\cZ_{G,q}) =
\fp^{-1} + \fZ[\cZ_{G/e,q} \cap \cZ_{G\smallsetminus e,q}]
- \fp^{-1} \fZ[\cZ_{G/e,q}] . $$

\smallskip

Similarly, one can define another probabilistic counting function, which
we denote here by $\fZ^\vee[f_1,\ldots,f_k]=1-\fZ[f_1,\ldots, f_k]$. This
is the probability of {\em not} having a common zero of the $f_i$. Just
as $\fZ[f_1,\ldots, f_k](\fq)$ is related to the number of points $N_\fq(X_\fp)$
by $\fZ[f_1,\ldots, f_k](\fq) = \fq^{-m} N_\fq (X_\fp)$, the probabilistic
counting function $\fZ^\vee[f_1,\ldots,f_k]$ satisfies
\begin{equation}\label{Zvee}
\fZ^\vee[f_1,\ldots,f_k](\fq)= \fq^{-m} (\fq^m-N_\fq (X_\fp))= 
\fq^{-m} N_{\fq} (\A^m \smallsetminus X_\fp).
\end{equation}
In other words, $\fZ^\vee[f_1,\ldots,f_k]$ is the probabilistic counting function
of the variety's complement. Thus, from \eqref{K0delcon} we have
\begin{equation}\label{Zveedelcon}
 \fZ^\vee[Z_{G,q}] =  \fZ^\vee[Z_{G\smallsetminus,q}, Z_{G/e,q}]   -\fq^{-1}  \fZ^\vee[Z_{G/e,q}].
\end{equation}

\medskip
\subsection{The case of the graph $K4$}

We now focus, in particular, on the polynomial countability question for
the case of the tetrahedron graph $K4$, and we show that it exhibits a more serious
failure of polynomial countability, which is not only due to the primes that
divide $q$.

\medskip

\begin{lem}\label{Znopol}
$\fZ[x^2 + 2x + 2]$ is not a polynomial.
\end{lem}

\proof First observe that there are infinitely many primes of the form $\fp=4k + 3$.
This follows from Dirichlet's theorem, showing more generally that there are
infinitely many primes in any arithmetic progression. (For this particular case
there is also a direct elementary proof, which we do not report here.)
Then notice that, for all primes $\fp$ of the form $\fp=4k + 3$, the polynomial 
$x^2 + 2x + 2$ has no solutions over $\F_\fp$. In fact, 
by the quadratic formula, $x^2 + 2x + 2$ has a solution if and only if $x = \frac{-2 \pm \sqrt{4 - 8}}{2} = -1 \pm \sqrt{-1}$.  This has a solution iff $\sqrt{-1} \in \mathbb{F}_\fp$.  Assume this is true, and that $\fp = 4k + 3$.  Let $a = \sqrt{-1}$, then we have $a^{\fp-1} = a^{4k + 2} = (a^2)^{2k + 1} = -1^{2k + 1} = -1$, but by Fermat's Little Theorem, $a^{\fp - 1} \equiv 1$.  
Therefore, $x^2 + 2x + 2$ has no solution for all primes of the form $\fp=4k + 3$.
Now, if $\fZ[x^2 + 2x + 2]$ were a polynomial, 
then it would have infinitely many zeros, since there are  infinitely many primes
$\fp$ (those of the form $\fp=4k+3$), where $x^2 + 2x + 2$ has no solution, but
$\fZ[x^2 + 2x + 2]$ is not identically zero: 
for instance, when $\fp = 5$, there are two roots, $x = 1$ and $x = 2$. 
Thus, $\fZ[x^2 + 2x + 2]$ is not polynomial.
\endproof

Consider then the graph $K4$ and in the case with $q=2$ (Ising model).
A direct calculation based on a version of the algorithm of Stembridge gives
\begin{equation}\label{NpK4}
 \begin{array}{ll}
N_{\fp}(\cZ_{K4,q=2}) = & \fp^5 \fZ [P(x)] +2\fp^3 \fZ[2+2x_4+x_4^2] \\[2mm]
& +2\fp^3 \fZ[2+2x_5+x_5^2] 
 +\fp^5-\fp^4-3\fp^3 +13 \fp^2-\fp-1 \\[2mm]
&   -\fp^5 \fZ[2,P(x)] -2\fp^3 \fZ[2, 2+2x_4+x_4^2]-2\fp^3 \fZ[2, 2+2x_5+x_5^2] \\[2mm]
& +\fZ[2] (\fp^6 -\fp^5 +\fp^4 +3\fp^3 -13\fp^2 +\fp +1) ,
\end{array}
\end{equation}
where the polynomial $P(x)$ is of the form
$$ P(x) = 4x_2x_4x_5 +8x_2x_3x_4x_5 +4x_2x_3x_4 +4x_2x_3x_5 +4x_3x_4x_5 $$
$$+4x_3x_4 +4x_2x_5 +2x_3x_4^2 +2x_3^2x_4 +x_3^2x_4^2 +2x_2x_3^2x_4  $$
$$+x_2x_3^2x_4^2 +2x_2x_3x_4^2 +2x_3x_4^2x_5 +2x_3^2x_4x_5 +x_3^2x_4^2x_5 $$
$$+2x_2^2x_4x_5x_3 +2x_2x_4^2x_5x_3 +x_3^2x_4^2x_5x_2 +2x_3^2x_4x_5x_2 $$
$$ +2x_2^2x_5 +2x_2^2x_4x_5 +2x_2^2x_5x_3 +2x_2x_5^2 +x_2^2x_5^2 +x_2^2x_4x_5^2 $$
$$ +2x_2x_4x_5^2+x_2^2x_3x_5^2+2x_2x_3x_5^2+x_2^2x_4x_5^2x_3+2x_2x_4x_5^2x_3. $$
The $\fZ[2]$ terms are only $1$ in the case where $2=0$, which we can ignore as we are looking
for non-trivial failures of polynomial countability. Among the remaining terms, we have seen in 
Lemma \ref{Znopol} that the $\fZ[2 + 2x + x^2]$ contribute non-polynomial expressions.
This strongly suggests that $\fZ[ Z_{K4,q} ]$ itself may be non-polynomial.
However, one needs to make sure that there are no cancellations (however unlikely) coming
from the term $\fp^5 \fZ [P(x)]$.

\smallskip

This is confirmed also by looking directly at the number of points for 
sufficiently many primes, as follows.

\begin{prop}\label{ZK4nonpol}
For $q=2$ (Ising model case), the function $N_{\fp}(\cZ_{K4,q=2})$ 
is non-polynomial.
\end{prop}

\proof The values for the first few primes $\fp\neq 2$ give the
following table.
\begin{center}
    \begin{tabular}{|c||c|c|c|c|c|c|c|c|}
    \hline
$\fp$ & 3 & 5 & 7 & 11 & 13 & 17 & 19 & 23 \\ \hline
$N_\fp$ & 413 & 4449 & 20901 & 180333 & 403025 & 1493449 & 2580541 & 6627909 \\ \hline
\end{tabular}
\end{center}

Proposition 2.2 of \cite{stembridge} shows that, since the graph $K_4$ has
six edges, if $N_{\fp}(\cZ_{K4,q=2})$ is given by a polynomial, then this polynomial would
have degree at most five. Thus, one obtains from the table above a system of
linear equations for the coefficients of this polynomial,
\begin{equation}\label{systemNp}
a_0 + \fp a_1  + \fp^2 a_2 + \fp^3 a_3 + \fp^4 a_4 + \fp^5 a_5  = N_{\fp} ,
\end{equation}
for $\fp$ and $N_{\fp}$ in the table above. Already when solving the
first five equations in this system, one finds only a solution with rational,
non-integer values of the $a_i$. By Proposition 6.1 of \cite{reine}, one
knows that if $N_{\fp}(\cZ_{K4,q=2})$ is polynomial with rational coefficients
then the coefficients must in fact be integers. Indeed, if one further
considers the remaining three equations from the table of values above,
one finds that there are no solutions. Thus, $N_{\fp}(\cZ_{K4,q=2})$ is
non-polynomial.
\endproof

While the result of Proposition \ref{ZK4nonpol} alone suffices to show that
the Potts model hypersurface of the graph $K4$ is not polynomially
countable, we have included the previous discussion to illustrate a possible 
explicitly source of non-polynomial terms.

\medskip
\subsection{Normalized Tutte polynomial}

We make a small additional remark on the trivial failures of polynomial
countability. 
In order to avoid the presence of an overall factor $q^{k(G)}$ in the multivariate
Tutte polynomial, one sometimes considers, instead of the polynomial \eqref{multiTutte},
the normalized version
\begin{equation}\label{normTutte}
\tilde Z_G (q,t)= q^{-k(G)} Z_G(q,t).
\end{equation}

The effect of this change on the counting function is only to alter it at
the primes $\fp$ that divide the number of states $q$. For example,
in the two simple examples of $C_4$ and $K_4$, one finds the following.

\begin{ex}\label{normZtreepoly}
For the normalized polynomial \eqref{normTutte}, the 
counting function for $C_4$ with $q=2$ is given by
$$ p^3-3 p^2 \fZ[2]+5 p^2+5 p \fZ[2]-7p-1, $$
which differs from the one for the Tutte polynomial \eqref{multiTutte}
by 
$$ \fZ[2] (-p^4 +p^3 + 2 p^2 -2 p - 1). $$
For the graph $K_4$ with $q=2$ the counting
function for the normalized case gives
$$ \begin{array}{c} p^5 \fZ[P(x)] +2 p^3\fZ[2+2 x_4+x_4^2]+2 p^3\fZ[x_5^2+2 x_5+2] \\[2mm]
+p^5-p^4 -3 p^3 +13 p^2 -p-1 +\fZ[2] (2 p^3 -15 p^2  +2 p), \end{array} $$
which again differs from the case of the original
Tutte polynomial only at $\fp=2$, with the difference given by
$$ \begin{array}{c} \fZ[2] (-p^6 +p^5 -p^4 -p^3 -2 p^2 +p -1 ) +p^5 \fZ[2,P(x)] \\[2mm]
+2p^3 \fZ[2, x_4^2+2 x_4+2]+2 p^3 \fZ[2, x_5^2+2x_5+2]. \end{array} $$
\end{ex}

\section{Thermodynamic averages and periods}

Coming back to the physical interpretation of the multivariate Tutte polynomial
$Z_G(q,t)$ as the partition function of the Potts model with $q$ spin states
on the graph $G$, the edge weights are of the form $t_e=\exp(\beta J_e) -1$, 
where $J_e\geq 0$ is the energy (in the ferromagnetic case) and $\beta$ is the
inverse temperature. One knows that, for a finite graph and in the
ferromagnetic case, there are no phase transitions, that is, no zeros of
$Z_G(q,t)$ that fall in the range $t_e\geq 0$ and the phase transitions
only appear in the limit $n\to \infty$ of a family of finite graphs $G_n$
approximating an infinite graph (see \cite{sokal}).

Given a function $f(J)=F(t)$ of the energies $J=(J_e)$, or of the edge weights $t=(t_e)$,
one can compute the thermodynamic average
\begin{equation}\label{thermoaverage}
\langle F \rangle = \frac{ \sum_{A\subseteq E} q^{k(A)} F(t_A) \prod_{e\in A} t_e }{
\sum_{A\subseteq E} q^{k(A)}   \prod_{e\in A} t_e }=\frac{1}{Z_G(q,t)}  \sum_{A\subseteq E} q^{k(A)} F(t_A) \prod_{e\in A} t_e ,
\end{equation}
where we write $F(t_A)=F(t)|_{t_e=0, \forall e\notin A}$. We are especially
interested here in the case where $F(t)$ is a polynomial function of the edge
variables, with rational coefficients. 

Moreover, we can further
average over a range of energies (at a fixed temperature), for example, by 
letting the edge variables range over the simplex $\Delta=\{ t=(t_e)\,|\, t_e\geq 0, \, 
\sum_e t_e = 1\}$. One then finds an expression that is formally very similar
to the parametric Feynman integral in perturbative quantum field theory, namely
\begin{equation}\label{thermoavint}
\frac{1}{Vol(\Delta)} \int_\Delta \langle F \rangle \, dv = \frac{1}{Vol(\Delta)} \int_\Delta
\frac{P_{G,F}(q,t)}{Z_G(q,t)} \, dv(t),
\end{equation}
where we write $P_{G,F}(q,t)$ for the polynomial
$$ P_{G,F}(q,t) = \sum_{A\subseteq E} q^{k(A)} F(t_A) \prod_{e\in A} t_e , $$
under the assumption that $F$ is itself a polynomial. 

\smallskip

The normalization factor in \eqref{thermoaverage}
can be computed easily: the volume of a regular $n$-dimensional
simplex with side length $a$ is given by the well known expression 
$$ Vol(\Delta_n(a))= \frac{a^n}{n!} \sqrt{\frac{n+1}{2^n}}. $$
Thus, the interesting number, whose nature one wants to investigate,
is the remaining expression
\begin{equation}\label{thermoavint2}
\int_\Delta \frac{P_{G,F}(q,t)}{Z_G(q,t)} \, dv(t). 
\end{equation}

\smallskip

We see that the expression \eqref{thermoavint2} that
we obtain in this way is formally similar to the parametric Feynman integral (although
with a very different physical interpretations),
with $Z_G(q,t)$ the multivariate Tutte polynomial instead of the graph polynomial,
and with $P_{G,F}(q,t)$ instead of the second Symanzik polynomial. 

\smallskip

An interesting difference with respect to the quantum field theory case, which
makes the Potts model case nicer, is that (as recalled above) the partition
function $Z_G(q,t)$, for a finite graph $G$ does not have any zeros in the
domain of integration. Thus, these integrals are genuine periods and are
not plagued by the infrared divergences problem that occurs in their
quantum field theoretic analogs, where the graph hypersurface intersects 
the simplex $\Delta$.

\smallskip

By analogy with the context of quantum field theory, one can then
ask the question of what kind of numbers one obtains by evaluating
the integrals \eqref{thermoavint2}, for arbitrary polynomials $F(t)$ with
rational coefficients.

\smallskip

The integral \eqref{thermoavint2} is a {\em period} of the variety
$\A^n \smallsetminus \cZ_{G,q}$, where $n=\# E(G)$: the integral of
an algebraic differential form over a domain of integration given by
a (semi)algebraic set defined by algebraic equations. Periods are a very interesting
class of numbers, see the detailed account given in \cite{kontza}.

\smallskip

In particular, what kind of numbers can arise as periods of a given variety depends
on the nature of the motive of the variety. For example, it was long conjectured and
recently proved in \cite{brown1} that the periods of mixed Tate motives over $\Z$
are $\Q[\frac{1}{2\pi i}]$-linear combinations of  multiple zeta values. 
The latter are expressions of the form
$$ \zeta(n_1,\ldots, n_r) = \sum_{0< k_1< \ldots < k_r} \frac{1}{k_1^{n_1} \cdots k_r^{n_r}}, $$
with integers $n_i\geq 1$ and $n_r \geq 2$.

\smallskip

\begin{figure}
\begin{center}
\includegraphics[scale=0.45]{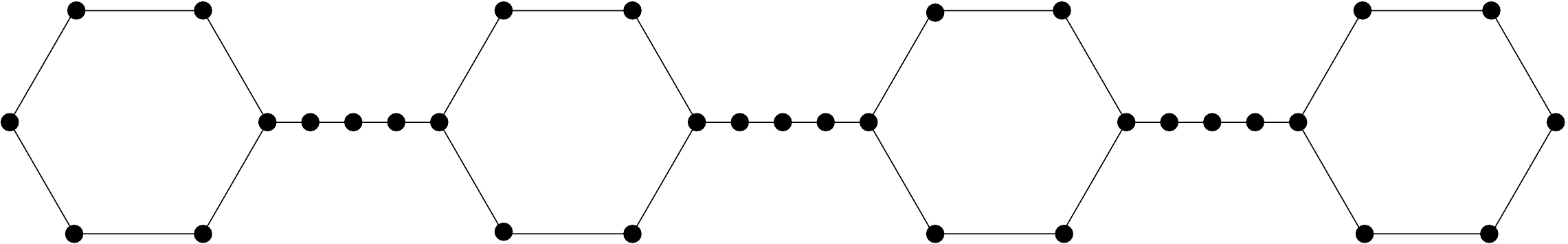}
\caption{A polygon polymer chain (from \cite{aluffi}). \label{FigPolyChain}}
\end{center}
\end{figure}

Thus, for example, in the cases of polygon polymer chains considered in
\cite{aluffi}, where it is shown by the explicit formula \eqref{polygonclass} 
that the motives involved are mixed Tate, one can conclude that the
periods \eqref{thermoavint2} must evaluate to combinations of
multiple zeta values, which is not obvious from the definition of the integral. 

We assume here some standard conjectures about motives, in particular
the Tate conjecture which implies that determining 
$N_{\fp}(X)$ for almost all primes $\fp$ determines the motive of $X$
(see \cite{andre}, \S 1.1).

\begin{prop}\label{MZV}
Let ${}^{(m,k)}G_N$ be a chain of linked polygons, obtained by joining
$N$ polygons, each with $m+1$ sides and connected by a chain of $k$ egdes,   
as illustrated in Figure \ref{FigPolyChain}. Then the thermodynamic averages
\eqref{thermoavint2} are $\Q[\frac{1}{2\pi i}]$-linear combinations of multiple zeta values.
\end{prop}

\proof The explicit formula given in Proposition 5.5 of
\cite{aluffi} for the class in the Grothendieck ring  gives
$$ \{ \cZ_{{}^{(m,k)}G_N,q} \}= \bT^{k(N-1)} \cdot \{ \cZ_{C_{m+1},q} \}, $$
where $ \{ \cZ_{C_{m+1},q} \} $ is as in \eqref{polygonclass}. This shows
that the class in the Grothendieck ring of the variety $\cZ_{{}^{(m,k)}G_N,q}$
is in the mixed Tate subring $\Z[\bL]\subset K_0(\cV_\Z)$. In principle, this
does not yet say that the motive of $\cZ_{{}^{(m,k)}G_N,q}$ is mixed Tate
(in the category of mixed motives), as the class $[\cZ_{{}^{(m,k)}G_N,q}]$
determines only its motivic Euler characteristic $\chi_{mot}(\cZ_{{}^{(m,k)}G_N,q})$.
However, if the conjecture holds, which predicts that knowledge of the counting function $N_{\fp}$
suffices to determine the motive, then we can conclude from the expression above
for the Grothendieck class that the motive is mixed Tate.
Then the main result of \cite{brown1} shows that the period \eqref{thermoavint2}
has to be a combination of multiple zeta values as stated.
\endproof

One can then ask a more precise question.

\begin{ques}\label{MZVques}
What combinations of multiple zeta values arise in the period computations
 \eqref{thermoavint2} for the polymer chain graphs ${}^{(m,k)}G_N$? How 
 do they behave as the parameters $m,k,N$ get large?
\end{ques}

\smallskip

With respect to this last question, and the more general question of
identifying criteria for when a graph satisfies the polynomial
countability condition, it is worth mentioning another important
difference between the quantum field theory case and the Potts
model hypersurfaces.

\smallskip

In the case of the graph hypersurfaces of quantum field theory, where the
hypersurface is defined by the vanishing of the Kirchhoff polynomial, one
can use the matrix-tree theorem to obtain some more precise information
on the periods. In fact, the Kirchhoff polynomial can be written as
$\Psi_G= \det(\cM_G)$, where $\cM_G$ is a matrix defined in terms
of the combinatorics (the incidence data) of the graph.  One then has
algebraic relations between determinants of minors of the matrix $\cM_G$,
which define relations between the corresponding 
Dodgson polynomials $\Psi_G^{I,J}$ of the graph. These identities
were used in \cite{brown2} to identify structural properties of graphs
whose associated Feynman period is a multiple zeta value. This
identifies a ``linear reducibility criterion" for graphs, which implies
the mixed-Tate property. Another setting in which the matrix-tree theorem
for Kirchhoff polynomial can be useful in the quantum field theory 
case is the result of \cite{aluffi5}, where one uses the determinant
description of the Kirchhoff polynomial to map the Feynman integral
computation to the complement of a determinant hypersurface.
In the case of the multivariate Tutte polynomials, however, it is well known
that one does not have an analog of the matrix-tree theorem. In fact, the
problem of computing the Tutte polynomials is known to be $\# P$-hard, see
\cite{jaeger}.

\begin{figure}
\begin{center}
\includegraphics[scale=0.9]{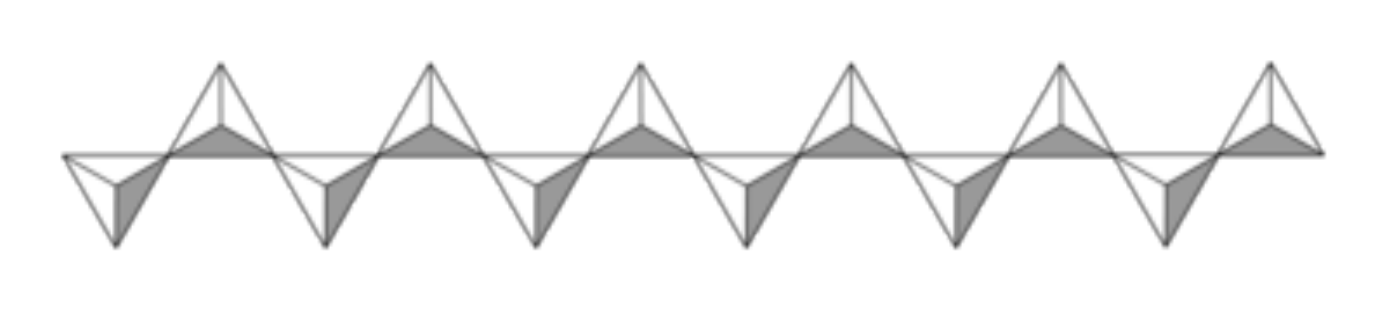}
\caption{Tetrahedra in a single-chain configuration. \label{FigChain}}
\end{center}
\end{figure}

\subsection{Potts models on tetrahedral chains}

Our previous observations on the tetrahedron graph suggest that the same
approach we just described for polygonal polymer chains
may lead to possibly very different answers when applied to a
different type of chain graphs, the tetrahedral chains that arise naturally in
the study of silicates.

\smallskip

The simplest example of Potts model, the Ising model with $q=2$ spin states,
has been extensively used in studying the properties of tetrahedral chains
that arise in silicates and Si-Al minerals (see \S 6 of \cite{putnis}). Example
of such Ising model analysis can be found in \cite{dove}, \cite{gemon}.

\smallskip

Here we consider the type of single chain tetrahedral configuration, as
illustrated in Figure \ref{FigChain}. This is realized, for instance, by
inosilicates  such as $({\rm Si} \, {\rm O}_3^{2-})_n$.

\smallskip

Recall that, for a graph $G$ that is obtained as a chain of graphs $G_1, \ldots, G_N$,
where $G_i$ and $G_{i+1}$ have a single vertex in common, the classes in
the Grothendieck ring of the corresponding Potts model hypersurface complements (with
$q\neq 0$) is simply a product (see Lemma 3.5 of \cite{aluffi})
\begin{equation}\label{prodZG}
\{ \cZ_{G,q} \} = \prod_{i=1}^N \{ \cZ_{G_i,q} \}.
\end{equation}

\smallskip

Thus, if $G_N$ is the graph obtained by linking together $N$ copies of
the tetrahedron graph $K4$, as in Figure \ref{FigChain}, the corresponding
class for $q\neq 0$ is given by $\{ \cZ_{G_N,q} \} = \{ \cZ_{K4,q} \}^N$,
or equivalently
$$ [  \cZ_{G_N,q} ] = \bL^{6N} - ( \bL^6- [\cZ_{K4,q}] )^N, $$
which implies that the counting function satisfies
$$ N_{\fp}(\cZ_{G_N,q}) = \fp^{6N} - (\fp^6 - N_{\fp}(\cZ_{K4,q}) )^N, $$
at primes $\fp$ that do not divide $q$. In the case of the Ising model, where $q=2$,
the counting function $N_{\fp}(\cZ_{K4,q=2})$ is the one computed in \eqref{NpK4}.

\smallskip

Thus, one sees that, unlike the case of the polymer chains analyzed in \cite{aluffi},
that involved the polynomial counting functions of polygons, here the counting
function involves non-polynomial expressions, as shown in Proposition \ref{ZK4nonpol}.
Although this does not prove that the motive is non-mixed Tate, as there could be
cancellations coming from the other terms in \eqref{NpK4}, it appears to be a 
good candidate to test. So one can ask the following question.

\begin{ques}\label{tetrahedraMZV}
Let $G_N$ be the graph obtained by linking together $N$ tetrahedra as in
Figure \ref{FigChain}. Are there choices of polynomials $F(t)$ with rational 
coefficients, for which the period \eqref{thermoavint2}, computed for this
graph is not a $\Q[\frac{1}{2\pi i}]$-linear combinations of multiple zeta values?
What kind of period is it, if that is the case?
\end{ques}

Notice that the analogous question in the quantum field theory setting
is not yet solved, even though there are explicit non-mixed Tate examples
of graph hypersurfaces as in \cite{brownschnetz}. In fact, there is not
yet an explicit example of graph hypersurface for which the Feynman
amplitude itself would be a non-mixed Tate period, although there are
good reasons to expect that, as explained in \cite{brownschnetz}.


\bigskip

{\bf Acknowledgment.} This paper is based on the results of  
the second author's summer research project, supported by the 
Summer Undergraduate Research Fellowship program at Caltech. 
The first author is partly supported by 
NSF grants DMS-0901221 and DMS-1007207.
The authors thank Paolo Aluffi and Bill Dubuque for useful conversations.

\end{document}